\documentstyle[11pt,amssymb,epsf]{article}
\def\crampest{\medmuskip = 1mu plus 1mu minus 1mu}
\def\uncramp{\medmuskip = 4mu plus 2mu minus 4mu}
\def\ben{\begin{equation}}
\def\een{\end{equation}}

  \let\n=\nu

\let\C=\Chi

\def\nn{\nonumber} \def\bd{\begin{document}} \def\ed{\end{document}}
\def\ds{\documentstyle} \let\fr=\frac \let\bl=\bigl \let\br=\bigr
\let\Br=\Bigr \let\Bl=\Bigl
\let\bm=\bibitem
\let\na=\nabla
\let\pa=\partial \let\ov=\overline
\newcommand{\be}{\begin{equation}}
\newcommand{\ee}{\end{equation}}
\def\ba{\begin{array}}
\def\ea{\end{array}}
\def\ft#1#2{{\textstyle{{\scriptstyle #1}\over {\scriptstyle #2}}}}
\def\fft#1#2{{#1 \over #2}}
\def\del{\partial}
\def\vp{\varphi}
\def\sst#1{{\scriptscriptstyle #1}}
\def\oneone{\rlap 1\mkern4mu{\rm l}}
\def\td{\tilde}
\def\wtd{\widetilde}
\def\ie{\rm i.e.\ }
\def\dalemb#1#2{{\vbox{\hrule height .#2pt
        \hbox{\vrule width.#2pt height#1pt \kern#1pt
                \vrule width.#2pt}
        \hrule height.#2pt}}}
\def\square{\mathord{\dalemb{6.8}{7}\hbox{\hskip1pt}}}
\newcommand{\ho}[1]{$\, ^{#1}$}
\newcommand{\hoch}[1]{$\, ^{#1}$}
\newcommand{\bea}{\begin{eqnarray}}
\newcommand{\eea}{\end{eqnarray}}
\newcommand{\ra}{\rightarrow}
\newcommand{\lra}{\longrightarrow}
\newcommand{\Lra}{\Leftrightarrow}
\newcommand{\ap}{\alpha^\prime}
\newcommand{\bp}{\tilde \beta^\prime}
\newcommand{\tr}{{\rm tr} }
\newcommand{\Tr}{{\rm Tr} }
\def\0{{\sst{(0)}}}
\def\1{{\sst{(1)}}}
\def\2{{\sst{(2)}}}
\def\3{{\sst{(3)}}}
\def\4{{\sst{(4)}}}
\def\5{{\sst{(5)}}}
\def\6{{\sst{(6)}}}
\def\7{{\sst{(7)}}}
\def\8{{\sst{(8)}}}
\def\n{{\sst{(n)}}}
\def\cA{{{\cal A}}}
\def\cF{{{\cal F}}}
\def\tV{\widetilde V}
\def\tW{\widetilde W}
\def\tH{\widetilde H}
\def\tE{\widetilde E}
\def\tF{\widetilde F}
\def\tA{\widetilde A}
\def\im{{{\rm i}}}
\def\tY{{{\wtd Y}}}
\def\ep{{\epsilon}}
\def\vep{{\varepsilon}}
\def\R{\rlap{\rm I}\mkern3mu{\rm R}}
\def\bD{{{\bar D}}}

\def\R{\rlap{\rm I}\mkern3mu{\rm R}}
\def\bD{{{\bar D}}}
\def\R{{{\Bbb R}}}
\def\C{{{\Bbb C}}}
\def\H{{{\Bbb H}}}
\def\CP{{{\Bbb C}{\Bbb P}}}
\def\RP{{{\Bbb R}{\Bbb P}}}
\def\Z{{{\Bbb Z}}}
\def\bA{{{\Bbb A}}}
\def\bB{{{\Bbb B}}}
\def\bC{{{\Bbb C}}}
\def\bR{{{\Bbb R}}}
\def\bD{{{\Bbb D}}}
\def\bE{{{\Bbb E}}}
\def\bZ{{{\Bbb Z}}}
\def\Re{{{\frak{Re}}}}
\def\Im{{{\frak{Im}}}}
\def\cosec{{\,\hbox{cosec}\,}}
\def\Gm{{\Gamma_{\!\! -}}}
\def\Gp{{\Gamma_{\!\! +}}}
\def\stan{{standard }}
\def\nonstan{{supernumerary }}

\def\cosech{{\hbox{cosech}}}

\def\etcyc{{\hbox{and cyclic}}}
\def\btheta{{\bar\theta}}

\def\vf{{\varphi}}
\def\hf{{\hat\phi}}
\def\fh{{\hat f}}
\def\ah{{\hat a}}
\def\eq#1{(\ref{#1})}
\newcommand{\w}[1]{\\[0.#1cm]}
\def\eb{ {\bar\epsilon} }
\def\cA{{\cal A}}
\def\cB{{\cal B}}
\def\cF{{\cal F}}
\def\cV{{\cal V}}
\def\cG{{\cal G}}
\def\cP{{\cal P}}
\def\cQ{{\cal Q}}
\def\hL{{\hat L}}
\def\ua{{\underline{\alpha}}}
\def\ub{{\underline{\beta}}}
\def\uc{{\underline{\gamma}}}

\newcommand{\undertilde}[1]{\underset{\widetilde{}}{#1}}

\newcommand{\tamphys}{\it George P. and Cynthia W. Mitchell  Institute
for Fundamental Physics,\\
Texas A\&M University, College Station, TX 77843, USA}

\newcommand{\auth}{{\Large
Jianwei Mei and C.N. Pope
}}

\begin{document}
\begin{flushright}
\hfill{
MIFP-07-22}\\
%\hfill{
%\bf hep-th/yymmnnn}
\end{flushright}

\begin{center}

{\Large {\bf New Rotating Non-Extremal Black Holes in $D=5$
              Maximal Gauged Supergravity
}}

\vspace{25pt}

\auth

\vspace{25pt}
\tamphys

\vspace{25pt}

\underline{ABSTRACT}

\end{center}

   We obtain new non-extremal rotating black hole solutions in
maximal five-dimensional gauged supergravity.  They are characterised by
five parameters, associated with the mass, the two angular momenta,
and two independently-specifiable charge parameters.  Two of the three
charges associated with the $U(1)^3$ Cartan subgroup of the $SO(6)$ gauge
group are equal, whilst the third can be independently specified.  These
new solutions generalise all the previously-known rotating solutions
in five-dimensional gauged supergravity with independent angular momenta.
They describe regular black holes, provided the parameters lie in appropriate
ranges so that naked singularities and closed-timelike curves (CTCs) are
avoided.  We also construct the BPS limit, and show that regular supersymmetric
black holes or topological solitons arise if the parameters are further
restricted in an appropriate manner.

\vspace{15pt}

\pagebreak

%\setcounter{page}{1}

%\tableofcontents

%\addtocontents{toc}{\protect\setcounter{tocdepth}{2}}

\newpage

%\tableofcontents

\newpage

%%%%%%%%%%%%%%%%%%%%%%%%%%%%%%%%%%%%%%%%%%%%%%%%%%%%%%%%

\section{Introduction}

%%%%%%%%%%%%%%%%%%%%%%%%%%%%%%%%%%%%%%%%%%%%%%%%%%%%%%%%

   The development of the AdS/CFT correspondence in string theory
\cite{adscft1,adscft2,adscft3} has led
to a growing interest in the construction of five-dimensional solutions in
gauged supergravity, which can be related to four-diemensional
boundary field theories.  Of particular interest in this context are the
solutions for five-dimensional black holes which are asymptotic to AdS
spacetime at large distance.  The AdS/CFT correspondence is more solidly
grounded in the case of BPS configurations, which are protected by
supersymmetry.  Asymptotically-AdS BPS black holes must
necessarily have non-zero rotation in order to be free from naked singularities
or other pathologies.  Thus when considering non-BPS black holes, it is
appropriate to include rotation as well, so that one can take a smooth
limit, free of pathologies, to reach the BPS configurations.

   From the standpoint of the AdS/CFT correspondence one is most interested
in finding such black-hole solutions within the maximal $SO(6)$-gauged
${\cal N}=8$ five-dimensional supergravity, since this is the theory that
arises from
the Pauli reduction of the type IIB superstring on $S^5$.  Black holes
with Abelian gauge fields can therefore carry 3 independent charges,
associated with the three $U(1)$ factors in the Cartan subgroup of
$SO(6)$.  Equivalently, one can think of such charged black holes as
solutions of ${\cal N}=2$ gauged five-dimensional supergravity coupled
to two additional vector multiplets.  (These two vectors, plus the
graviphoton of the ${\cal N}=2$ supergravity itself, carry the three
charges.)  The general black-hole solution should then be characterised
by its mass, the two independent angular momenta associated with
rotations in the two orthogonal spatial 2-planes, and the three independent
charge parameters.

   The currently-known charged non-extremal rotating black-hole solutions in
the five-dimensional gauged supergravity are as follows.  For black holes
with two the independent rotation parameters, the uncharged solution (Kerr-AdS)
was found in \cite{hawhuntay} and the solution with all three charges equal
was found in \cite{cclp1}.  In addition, a solution with only one charge
non-zero was found in \cite{cclp2}, and a solution where two charges are
equal, with the third having a specific non-vanishing charge related to the
other two was found in \cite{cclp3}.  In the much simpler situation
where the two rotation parameters are set equal, the solution with three
independent charges was obtained in \cite{cvlupo}.

    The general non-extremal solution with two independent rotations and
three independent
charges is still unknown.  The purpose of the present paper is to advance
one step further to this goal.  Here, we construct the solution for
a non-extremal black hole in five-diemensional gayged supergravity with
the two independent angular momenta and with {\it two} independent charge
parameters.  This corresponds to the situation where two of the
three charges in the general solution are set equal, whilst the third can
be independently specified.  For appropriate specialisations of the
charge parameters in our new solution, all the previous cases in \cite{cclp1},
\cite{cclp2} and \cite{cclp3} mentioned above can be obtained.

   Having constructed the non-extremal solution we may also consider the
limit where a BPS bound is attained.  In general this describes a
supersymmetric configuration with singularities or closed timelike curves
(CTCs) outside a Killing horizon.  By making a further specialisation of the
parameters, we can obtain a class of ``regular'' BPS black holes, with
neither naked singularities nor naked CTCs.  This additional specialisation
also ensures that the Hawking temperature at the horizon is zero, as it
must be for a regular supersymmetric black hole.

\section{The Black Hole Solution}

   As with earlier work on non-extremal asymptotically-AdS rotating black
holes, the complexity of the equations of motion and the absence of any
solution-generating techniques means that the only practicable method
for finding solutions involves a large measure of guesswork and trial and
error, followed by explicit verification of the field equations.  In this
task we were aided greatly by knowledge of the solutions in the
previously-obtained special cases, especially those in  \cite{cclp1},
\cite{cclp2} and \cite{cclp3}.

   The five-dimensional Lagrangian for the bosonic sector of ${\cal N}=2$
gauged supergravity coupled to two vector multiplets can be written as
%%%%%
\be
L=\sqrt{-g}\left[R -\ft12\sum_{\alpha=1}^2(\del\varphi_\alpha)^2
  +\sum_{i=1}^3\left( 4g^2X_i^{-1} -\ft14  \cF^i_{\mu\nu}  \cF^{i\mu\nu}
  \right)
\right]+\frac{1}{24}|\varepsilon_{ijk}|\varepsilon^{uv\rho\sigma\lambda}
 \cF_{uv}^i  \cF_{\rho\sigma}^j  \cA_\lambda^k\,, \ee
%%%%%
where $g$ is the gauge-coupling constant, and the quantities $X_i$ are formed
from the two scalar fields $\varphi_1$ and $\varphi_2$ in the vector
multiplets:
%%%%%
\be
X_1=e^{-\ft1{\sqrt6} \varphi_1 -\ft1{\sqrt2}\varphi_2}\,,\qquad
X_2=e^{-\ft1{\sqrt6} \varphi_1 +\ft1{\sqrt2}\varphi_2}\,,\qquad
X_3= e^{\ft2{\sqrt6}\varphi_1}\,.
\ee
%%%%%
We find that the following is a solution of the resulting equations of motion:
%%%%%
\crampest
\bea 
ds^2\!\!\!&=& \!\!\! H_1^{2/3} H_3^{1/3} \left\{
(x^2 -  y^2)\Big(\fft{dx^2}{X}  -  \fft{dy^2}{Y}
 \Big) - \fft{x^2 X (dt + y^2 d\sigma)^2}{(x^2-y^2) f H_1^2}   + 
   \fft{y^2 Y\left[dt + (x^2+2m s_1^2) d\sigma\right]^2}{(x^2-y^2)(\gamma +y^2)
   H_1^2}\right.
\nn\\
&&\quad \qquad \qquad 
   \left.- U\left(dt + y^2 d\sigma + \fft{(x^2-y^2) f H_1 \left[ab
d\sigma +(\gamma+y^2)d\chi\right]}{ab(x^2-y^2) H_3 -2m s_3 c_3
(\gamma+y^2)}\right)^2
\right\}\,,\label{metric1}\\
 \cA^1 &=&
  \cA^2= \fft{2ms_1 c_1 (dt + y^2d\sigma)}{(x^2-y^2) H_1}\,,\nn\\
 \cA^3 &=&\fft{2m\left\{ s_3 c_3(dt+y^2 d\sigma)-(s_1^2-s_3^2)
\left[ab d\sigma +(\gamma+y^2) d\chi\right]\right\} }{(x^2-y^2)
H_3}\,,
\label{Afields}\\
X_1 &=& X_2 = \Big(\fft{H_3}{H_1}\Big)^{1/3}\,,\qquad
  X_3 =\Big(\fft{H_1}{H_3}\Big)^{2/3}\,,\label{scalars}
\eea
\uncramp
%%%%%
where
%%%%%
\bea
f&=& x^2 + \gamma + 2 m s_3^2\,,\qquad \gamma = 2ab s_3 c_3 + (a^2+b^2) s_3^2
\,,\nn\\
U&=& \fft{\left[ab (x^2-y^2)H_3 - 2m s_3 c_3
(\gamma+y^2)\right]^2}{
  (x^2-y^2)^2(\gamma+y^2) f H_1^2 H_3}\,,\nn\\
H_1 &=& 1 + \fft{2m s_1^2}{x^2-y^2}\,,\qquad
H_3 = 1 + \fft{2m s_3^2}{x^2-y^2}\,,
\eea
%%%%%
and $s_i\equiv \sinh\delta_i$, $c_i\equiv \cosh\delta_i$.  The
functions $X$ and $Y$ are given by
%%%%%
\crampest
\bea
X &=& \fft{-2m x^2 + (\td a^2 +x^2)(\td b^2+x^2)+ g^2 (\td a^2 + 2m s_1^2 +x^2)
  (\td b^2 + 2m s_1^2 +x^2)(2m s_3^2 + \gamma + x^2)}{x^2}\,,\nn\\
Y &=& \fft{(\td a^2+y^2)(\td b^2 +
y^2)\left[1+g^2(\gamma+y^2)\right]}{y^2}\,, \ {\rm with}\ \ \td
a\equiv a c_3 + b s_3\ ,\ \td b\equiv b c_3 + a s_3\,.
\eea
\uncramp
%%%%%
%where
%%%%%
%\be
%\td a\equiv a c_3 + b s_3\,,\qquad \td b\equiv b c_3 + a s_3\,.
%\ee
%%%%%
The solution is characterised by the mass parameter $m$, the two
rotation parameters $a$ and $b$, and the two charge parameters
$\delta_1$ and $\delta_3$.  It is evident from (\ref{Afields})
that the charges carried by the gauge fields $ \cA^1$ and $ \cA^2$
are equal, whilst that carried by $ \cA^3$ is independently
specificiable.

   The solution can be rewritten in an asymptotically non-rotating frame,
in terms of a canonically-normalised time coordinate $\tau$ and azimuthal
coordinates $\phi$ and $\psi$ having independent periodicities $2\pi$ by means
of the transformation
%%%%%
\bea
t&=& \fft{(1+g^2\gamma) \tau}{\Xi_a \Xi_b}-
         \fft{a(a^2+\gamma)\phi}{(a^2-b^2)\Xi_a} +
  \fft{b(b^2+\gamma)\psi}{(a^2-b^2)\Xi_b}\,,\nn\\
\sigma &=& \fft{g^2 \tau}{\Xi_a\Xi_b} -\fft{a \phi}{(a^2-b^2)\Xi_a} +
       \fft{b \psi}{(a^2-b^2)\Xi_b} \,,\nn\\
\chi &=& \fft{g^4 a b\tau}{\Xi_a\Xi_b} -
\fft{b\phi}{(a^2-b^2)\Xi_a}
   +\fft{a \psi}{(a^2-b^2) \Xi_b}\,,\label{azdef}
\eea
%%%%%
where
%%%%%
\be
\Xi_a \equiv 1-g^2 a^2\,,\qquad \Xi_b \equiv 1-g^2 b^2\,.
\ee
%%%%%
It is also useful to defined new coordinates $r$ and $\theta$ to
replace $x$ and $y$,
%%%%%
\bea x^2&=& r^2 - \gamma -\ft23 m(2s_1^2 + s_3^2)\ ,\nn\\
   y^2 &=& -\td a^2 \cos^2\theta -\td b^2 \sin^2\theta
    = -\gamma -a^2 \cos^2\theta -b^2 \sin^2\theta\ .\label{latdef}
\eea
%%%%%
For later convenience, we also define a new radial metric function
$\Delta_r(r)$ by
%%%%%
\be \Delta_r(r)=\frac{x^2X(x)}{r^2}\ \ \ \ \Longrightarrow\ \ \
\fft{dx^2}{X(x)} = \fft{dr^2}{\Delta_r(r)}\,, \ee
%%%%%
where $x$ is given in (\ref{latdef}). After rewriting the full
metric (\ref{metric1}) in terms of these new coordinates as
defined in (\ref{azdef}) and (\ref{latdef}), it can be seen that
it describes a rotating black hole with an horizon of $S^3$
topology located at the largest root $r=r_0$ of the function
$\Delta_r(r)$ . At large distance, $r\rightarrow\infty$ , the
metric approaches anti-de Sitter spacetime $\left(R_{\mu\nu}
\rightarrow -4 g^2 g_{\mu\nu} \right)$ ,
%%%%%
\be ds^2 \sim -\fft{(1+g^2 r^2) \Delta_\theta}{\Xi_a\Xi_b} d\tau^2
+
   \fft{dr^2}{g^2 r^2} + \fft{\rho^2 d\theta^2}{\Delta_\theta}
    + \fft{r^2+a^2}{\Xi_a} \sin^2\theta d\phi^2 +
   \fft{r^2+b^2}{\Xi_b} \cos^2\theta d\psi^2\,,\label{asmet}
\ee
%%%%%
where
%%%%%
\be
\rho^2\equiv r^2 + a^2\cos^2\theta + b^2 \sin^2\theta\,,\qquad
  \Delta_\theta \equiv 1- g^2 a^2 \cos^2\theta - g^2 b^2 \sin^2\theta\,.
\ee
%%%%%

\section{Conserved Charges and Thermodynamics}

   The angular momenta can be determined from the Komar integrals
$J=1/(16\pi) \int_{S^3} {*dK}$, where $K=K_\mu dx^\mu$ and
$K^\mu \del/\del x^\mu = \del/\del\phi$ or $\del/\del\psi$.  These give,
respectively,
%%%%%
\bea J_\phi &=& \frac{\pi m\left[a\left(c_3^2+s_3^2 + \Xi_b
(s_1^2-s_3^2)
\right) + b c_3s_3(1+g^2 a^2) \right]}{2\Xi_a^2 \,\Xi_b }\,,\nn\\
J_\psi &=& \frac{\pi m\left[b\left(c_3^2+s_3^2 + \Xi_a
(s_1^2-s_3^2) \right)+ a c_3s_3(1+g^2 b^2)  \right]}{2\Xi_a\Xi_b^2
}\,, \eea
%%%%%

   The conserved electric charges are given by
$Q_i = 1/(16\pi) \int_{S^3} (X_i^{-2} {* \cF^i} +\ft12
|\varepsilon_{ijk}|\,  \cA^j\wedge  \cF^k)$, evaluated over the
sphere at infinity.  We find
%%%%%
\be Q_1= Q_2 = \fft{\pi m s_1 c_1}{2\Xi_a\, \Xi_b}\,,\qquad Q_3 =
\fft{\pi m[s_3 c_3 - g^2ab(s_1^2-s_3^2)]}{2\Xi_a\, \Xi_b}\,. \ee
%%%%%

    The conserved mass $E$ could in principle be calculated using the conformal
technique of Ashtekar, Magnon and Das \cite{ashmag,ashdas}, but in practice it
is easier to evaluate it by integrating the first law of thermodynamics,
%%%%%
\be
dE= T dS + \Omega_\phi dJ_\phi  + \Omega_\psi dJ_\psi + \sum_i \Phi_i dQ_i\,,
\label{firstlaw}
\ee
%%%%%
where $T$ is the Hawking temperature, $S$ is the entropy,
$\Omega_\phi$ and $\Omega_\psi$ are the angular velocities of the
horizon and $\Phi_i$ are the potential differences between the horizon and
infinity.  To do this, we first construct the Killing vector $\ell$ that
becomes null on the horizon at $r=r_0$, given by
%%%%%
\be
\ell= \fft{\del}{\del \tau} + \Omega_\phi\, \fft{\del}{\del\phi}
   + \Omega_\psi\, \fft{\del}{\del\psi}\,.\label{elldef}
\ee
%%%%%
We find that the anglar velocities are given by
%%%%%
\bea \Omega_\phi&=&\fft{b(ab+2ms_3 c_3)+a[1+g^2(b^2+r_0^2 +
        \ft23 m(s_1^2-s_3^2))][r_0^2-\ft43 m(s_1^2-s_3^2)]}{
  ab(ab+2m s_3 c_3)+ [a^2+b^2+r_0^2+\ft23m(s_1^2-s_3^2)]
          [r_0^2-\ft43 m(s_1^2-s_3^2)]}\,,\nn\\
\Omega_\psi&=& \fft{a(ab+2ms_3 c_3)+b[1+g^2(a^2+r_0^2 +
        \ft23 m(s_1^2-s_3^2))][r_0^2-\ft43 m(s_1^2-s_3^2)]}{
  ab(ab+2m s_3 c_3)+ [a^2+b^2+r_0^2+\ft23m(s_1^2-s_3^2)]
          [r_0^2-\ft43 m(s_1^2-s_3^2)]}\,,\label{angvel} \eea
%%%%%

   The surface gravity $\kappa$ is given by
%%%%%
\be
\kappa^2 = \lim_{r\rightarrow r_0} \fft{g^{\mu\nu}(\del_\mu \ell^2)
          (\del_\nu\ell^2)}{(-4\ell^2)}\,,
\ee
%%%%%
 From this, we find that the Hawking temperature $T=\kappa/(2\pi)$ is
given by
%%%%%
\be T=\fft{r_0 \Delta'_r(r_0) \sqrt{r_0^2-\ft43 m(s_1^2-s_3^2)} }{
          4\pi\{ ab(ab+2m s_3 c_3)+ [a^2+b^2+r_0^2+\ft23m(s_1^2-s_3^2)]
          [r_0^2-\ft43 m(s_1^2-s_3^2)]\} }\,,\label{hawkingT}
\ee
%%%%%
where $\Delta'_r(r_0)$ means the derivative of $\Delta_r(r)$
evaluated at $r=r_0$ .

   The entropy $S$ is equal to a quarter of the area of the
3-sphere horizon at $r=r_0$, and is given by
%%%%%
\be
S= \fft{\pi^2[ab(ab+2m s_3 c_3)+
       (a^2+b^2+r_0^2 +\ft23 m (s_1^2-s_3^2))(r_0^2-\ft43 m
   (s_1^2-s_3^2))]}{2\Xi_a\, \Xi_b\,
          \sqrt{r_0^2-\ft43m (s_1^2-s_3^2)}}\,.
\ee
%%%%%

   Finally, the electrostatic potentials $\Phi_i$ on the horizon are given by
evaluating $\ell^\mu\,  \cA^i_\mu$ at $r=r_0$.  The potentials at
infinity vanish in the gauge we are using.  Thus we find
%%%%%
\bea
\Phi_1&=& \Phi_2 = \fft{2m s_1 c_1 [r_0^2 -\ft43 m (s_1^2-s_3^2)]}{
  ab(ab+2m s_3 c_3)+
       (a^2+b^2+r_0^2 +\ft23 m (s_1^2-s_3^2))(r_0^2-\ft43 m
   (s_1^2-s_3^2))}\,,\nn\\
\Phi_3 &=& \fft{2m[s_3 c_3 (r_0^2 + \ft23 m(s_1^2-s_3^2)) + ab(s_1^2-s_3^2)]}{
ab(ab+2m s_3 c_3)+
       (a^2+b^2+r_0^2 +\ft23 m (s_1^2-s_3^2))(r_0^2-\ft43 m
   (s_1^2-s_3^2))}\,.
\eea
%%%%%

   Using all the above results, we are in a position to evaluate the
right-hand side of the first law (\ref{firstlaw}), and to integrate it up
to obtain the conserved mass $E$.  It is highly non-trivial that the
right-hand side turns out to be an exact differential, and this provides
a useful check on the algebra.  We find that the conserved mass is given by
%%%%%
\bea E&=& \fft{m\pi \left[(c_3^2+s_3^2)(2\Xi_a + 2\Xi_b - \Xi_a\,
\Xi_b)+4 g^2 ab s_3 c_3 (\Xi_a+\Xi_b)\right]}{4 \Xi_a^2\, \Xi_b^2} \nn\\
&&+ \fft{m\pi (s_1^2-s_3^2)\left[ 2\left(\Xi_a + \Xi_b + g^4(a^4+b^4)\right)
  + g^2(a^2+b^2)(\Xi_a\, \Xi_b -2) \right]}{4 \Xi_a^2\, \Xi_b^2}\,.
\eea
%%%%%

  It is straightforward to check that the angular momenta, electric
charges and conserved mass, along with the
other thermodynamic quantities calculated here, agree in the appropriate limits
with previous results.  Thus setting $s_1=s_3$ yields the previous results
in \cite{cclp1} for the case of 3 equal charges; setting $s_1=0$ yields the
results in \cite{cclp2} and \cite{cclp3} for the case of a single
non-zero charge; and setting instead $s_3=0$ yields the results in
\cite{cclp3} for the case with 2 charges equal and the third non-vanishing
but related to the other two.

\section{BPS Limit and Supersymmetric Black Holes}

   A BPS limit of the non-extremal solutions will arise if the conserved
charges satisfy the condition\footnote{Equivalent BPS conditions arise for
all other choices of signs in this equation.}
%%%%%
\be
E = g J_\phi + g J_\psi +\sum_{i=1}^3 Q_i\,.\label{bps}
\ee
%%%%%
The solution then admits a Killing spinor, implying that it is a
supersymmetric supergravity background.  Substituting our results from the
previous section, we find (\ref{bps}) implies that
%%%%%
\be
e^{2\delta_1+2\delta_3} = 1 + \fft{2}{g(a+b)}\,.\label{bpscon}
\ee
%%%%%
(Recall that $\delta_1$ and $\delta_3$ are the charge parameters in the
original metric, with $s_i=\sinh\delta_i$, etc.)

   The existence of a Killing spinor $\eta$ allows one to write down an
everywhere-timelike Killing vector $K^\mu=\bar\eta \Gamma^\mu\eta$.  This
will take the form
%%%%%
\be K= \fft{\del}{\del\tau} + g\, \fft{\del}{\del\phi} + g\,
\fft{\del}{\del\psi}\,. \ee
%%%%%
Because its admits a spinorial square root, the Killing vector $K$ has a
manifestly negative norm (see, for example, \cite{cvgilupo}, and also
\cite{cclp1}), and in fact one can show that when (\ref{bpscon}) is
satisfied
%%%%%
\bea K^2 &=& -h_1^{-4/3}h_3^{-2/3}\, \left[\rho^2
+\frac{m\left[(2+g a+g b)^2-g^2(a+b)^2e^{4\delta_3}\right]
\left(3\Delta_\theta - (1+ga)(1+gb)\right)}{ 6g^2
(a+b)^2(1+ga)(1+gb)(2+ga+gb) e^{2\delta_3}}\right.\nn\\
&&\left.\ \ \ \ \ \ \ \ \ \ \ \ \ \ \ \ \ \ \ -\frac{2m}{3g^2
(a+b)^2e^{2\delta_3}}\right]^2\,, \label{Ksq} \eea
%%%%%
where
%%%%%
\be h_1=\rho^2 +\frac{2}{3}m(s_1^2-s_3^2)\ \ ,\ \ \  h_3= \rho^2
-\frac{4}{3}m(s_1^2-s_3^2)\,. \ee
%%%%%

  This result is useful for studying the occurrence of closed timelike curves
(CTCs) in the BPS metric.  First, we note that the metric can be cast in
the form
%%%%%
\bea
ds^2&=&-\frac{r^2\Delta_r(r)\Delta_\theta\sin^2\theta\cos^2\theta
dt^2}{ \Xi_a^2\, \Xi_b^2\,
B_{\phi}B_{\psi}}+h_1^{2/3}h_3^{1/3}\left[
\frac{d\theta^2}{\Delta_\theta}+\frac{dr^2}{\Delta_r(r)}\right]\nonumber\\
&&+B_{\psi}(d\psi+v_1d\phi+v_2dt)^2+B_{\phi}(d\phi+v_3dt)^2\,,
\label{ds2CTC} \eea
%%%%%
where the functions $B_\phi$, $B_\psi$ and $v_i$ can be read off by comparing
(\ref{ds2CTC}) with the original form of the metric.  In order not to have
CTCs, it must be that $B_\phi$ and $B_\psi$ are non-negative outside the
horizon.  After imposing
(\ref{bpscon}), we can write
%%%%%
\be K^2=
-\frac{r^2\Delta_r(r)\Delta_\theta\sin^2\theta\cos^2\theta}{
\Xi_a^2\, \Xi_b^2\, B_{\phi}B_{\psi}} + B_\psi(g+v_1 g +v_2)^2 +
            B_\phi (g+v_3)^2\,,
\ee
%%%%%
and so on the horizon, where $\Delta_r(r)=0$, the negativity of $K^2$ implies
that $B_\phi$ or $B_\psi$ must be negative, and hence except for special cases
there will be CTCs on and outside the horizon in the BPS solutions.

\subsection{Supersymmetric black holes}

   One way to avoid the occurrence of CTCs outside the horizon in the BPS
solutions is to arrange by means of a further condition on the parameters
that $K^2$, given by (\ref{Ksq}), actually vanishes on the horizon.  As in
cases studied previously, such as that of three equal charges in \cite{cclp1},
this condition is precisely equivalent to the condition that the derivative
of the metric function $\Delta_r(r)$ vanishes on the horizon at $r=r_0$.  In
other words, it has a double root there:
%%%%%
\be
\Delta_r(r_0)=0=\Delta'_r(r_0)\,.\label{delprime}
\ee
%%%%%
As can be seen from (\ref{hawkingT}), this means that the Hawking temperature
vanishes.  This is indeed a necessary condition for having a regular
supersymmetric black hole, since the inequivalent energy distribution functions
for bosons and fermions in a thermal state at non-zero temperature are
manifestly incompatible with supersymmetry.

   A convenient way to solve the zero-temperature condition
(\ref{delprime}) in addition to the BPS condition (\ref{bpscon}) is to
regard (\ref{bpscon}) as placing a constraint on
the value of the gauge-coupling constant $g$ as a function of the rotation
and charge parameters.  (This has the advantage of allowing not only the
two angular momenta, but also the two charge parameters, to be adjusted
freely, and this makes it easier to compare results with previously-known
cases such as $\delta_1=\delta_3$, $\delta_1=0$ or $\delta_3=0$.)
The zero-temperature condition (\ref{delprime})
can then be solved for the mass parameter $m$, implying that
%%%%%
\be M= \fft{e^{\delta_1+\delta_3}\left[
(a^2+b^2)\sinh(2\delta_1+2\delta_3) +
    2ab \cosh(2\delta_1+2\delta_3)\right]}{2\sinh(\delta_1+\delta_3)
              \sinh 2\delta_1}\,.\label{zeroT}
\ee
%%%%%

   If the solution is chosen so that both (\ref{bpscon}) and (\ref{zeroT})
are satisfied, then it can describe a regular supersymmetric black hole.
It is still necessary to restrict the remaining 3 parameters to lie within
appropriate regions, in order that the metric be free of any CTCs outside the
horizon, but these remaining conditions take the form of inequalities
rather than further functional relations between the parameters.  They
are generalisations of the restrictions found in \cite{cclp1} for the
case when the three charges were equal.  One can, for example, see that if
$ga$ and $gb$ are sufficiently small and positive, and the charge parameter
$\delta_3$ 
is sufficiently large, then there will be no CTCs outside the horizon.

   The supersymmetric black holes that we have obtained here will correspond
to the $Q_1=Q_2$ specialisation of the supersymmetric 3-charge black holes
coostructed in \cite{krlure}.

\subsection{Topological solitons}

    A second way of eliminating CTCs in the BPS solutions is if the
product $B_\phi B_\psi$ is proportional to $\Delta_r(r)$, and hence one or
other of $B_\phi$ or $B_\psi$ vanishes on the horizon.  In this case,
the BPS condition (\ref{bpscon}) is supplemented by the further condition
%%%%%
\be
m=\frac{2k_3k_4(a+b)(1+ga)(1+gb)(2+ga+gb)e^{2\delta_3}}{k_1^2k_2}\,,
\ee%
%%%%
with
%%%%%
\bea
k_1&=&(2+ga+gb)^2 -g^2(a+b)^2 e^{4\delta_3}\ ;\\
k_2&=&(2+ga+gb)(a+b+2gab)-(a+b)(2+ga +gb+2g^2ab)e^{4\delta_3}\ ;\nonumber\\
k_3&=&(2+ga+gb)\left[2a-gb^2+gab(1-ga-gb)\right]+g(a+b)^2(2+gb
+g^2ab)e^{4\delta_3}\ ;\nonumber\\
k_4&=&(2+ga+gb)\left[2b-ga^2+gab(1-ga-gb)\right]+g(a+b)^2(2+ga
+g^2ab)e^{4\delta_3}\nonumber\,.
\eea
%%%%%
(In this case, we have chosen to use (\ref{bpscon}) to eliminate
$\delta_1$.) The metric now describes a smooth topological
soliton, with $r=r_0$ being a regular origin of polar coordinates
at which $B_\phi\rightarrow 0$, and free of conical singularities,
provided that the quantisation condition
%%%%%
\be \frac{ak_4- b
k_3}{g(a-b)b(1-ga)}\left[\frac{2g^2b}{k_1}-\frac{1+gb}{
k_2}-\frac{g(a-b)(1-gb)}{k_4}\right]=1 \ee
%%%%%
is satisfied.  These topological solitons generalise examples found in
\cite{cclp1} in the case that the three charges were equal.

\section{Conclusions}

    The most
general non-extremal
black holes with an $S^3$ horizon topology in maximal $SO(6)$-gauged
five-dimensional supergravity would be
characterised by a total of six parameters, comprising the mass, the two
independent angular momenta, and three independent electric charges supported
by the three abelian gauge fields in the $U(1)^3$ Cartan subgroup of
$SO(6)$.  They could equivalently be regarded as solutions in ${\cal N}=2$
gauged supergravity coupled to two vector multiplets.

   In this paper, we
have constructed the most general such non-extremal rotating black holes
found to date.  They are characterised by five parameters, namely the mass,
the two angular momenta, and two independently-specifiable charge
parameters.  They correspond to the situation where two of the three
charges in the most general solution are set equal, but with no
restrictions otherwise.  These solutions encompass and extend
all previously-obtained
results for black holes with independent rotation parameters in
five-dimensional gauged supergravity.

    We calculated the conserved angular momenta and charges for the new
solutions; the entropy and Hawking temperature; and the angular velocities
and electric potentials on the horizon.  From this, we showed that the
first law of thermodynamics is integrable, and we obtained the expression
for the mass of the black holes.

   We then studied the BPS limit of the solutions, and showed how further
restrictions on the remaining parameters would give rise to regular
supersymmetric black holes and to smooth topological solitons.

   The results we have obtained in this paper should have applications in
the sudy of the AdS/CFT correspondence.  It would be of considerable interest
to find the more general 6-parameter black-hole solutions in
five-dimensional maximal gauged supergravity, in which the three electric
charges, as well as the mass and the two angular momenta,
are independently specificable.  These can be expected to be considerably 
more complicated than the solutions constructed until now.

\end{document}